# PARTICLE MASSES AND THE COSMOLOGICAL 'CONSTANT' IN FIVE DIMENSIONS

Paul S. Wesson

1. Department of Physics and Astronomy, University of Waterloo, Waterloo, Ontario N2L 3G1, Canada.

2. Herzberg Institute of Astrophysics, National Research Council, Victoria, B.C. V9E 2E7, Canada.

Abstract: I give metrics and equations of motion in 5D general relativity, and use the conservation of momentum and conformal transformations to study the possible variability of particle masses and the cosmological 'constant'. It is feasible that all particles are photon-like and travel on null paths in 5D, that massive particles are perturbations of the extra dimension which intrude into 4D, and that the cosmological 'constant' is a 5D / 4D gauge term. In the simplest case, a particle has a local cosmological 'constant' whose magnitude is proportional to the square of the mass.

Correspondence: mail to (1) above; email: psw.papers@yahoo.ca

1. <u>Introduction</u>

In higher-dimensional gravity, the equations of motion are expressed in $N(>4)$D, and generally show extra effects when applied to 4D spacetime. These effects provide in principle a way to test the kind of $N$D theory expected to yield a unified account of the interactions of physics. Yet despite its importance, there remains considerable confusion about this subject. The objective of the present account is clarification, particularly in regard to the value of the rest mass $m$ of a test particle and the value of the cosmological 'constant' $\Lambda$ that typifies the vacuum in which the particle moves. I will show that in 5D, there is a coordinate frame or gauge in which $m$ and $\Lambda$ have a simple relationship that is potentially testable.

Interest in the nature of $m$ and $\Lambda$ has a long history, of course, extending back to Mach and Einstein [1]. Nowadays these subjects are studied under the appellations of the hierarchy problem and the cosmological-'constant' problem. In Einstein's general relativity, both $m$ and $\Lambda$ are true constants, but in modern Kaluza-Klein theory with a non-compactified fifth dimension both may be variable. Current versions of 5D relativity include space-time-matter (STM) theory and membrane (M) theory. These are similar mathematically, but different physically [2, 3]. Both are in agreement with observations, and gain support from evidence for an accelerating universe [4]. This can be interpreted as due to a decaying cosmological 'constant', persistent inflation or a scalar field, all from 5D [5, 6]. The classical scalar field of 5D relativity may be related to the Higgs field of quantum field theory [7], and could be responsible for boosting the masses of particles from the inherently zero values predicted by supersymmetry to the finite values ac-tually observed. In this regard, it is now known that *null* geodesics in 5D correspond to the timelike geodesics of massive particles observed in 4D. This applies to both STM-theory and M-theory [8, 9], and means that the 5D and 4D intervals preserve conventional causality $\left(dS^2 = 0,\ ds^2 \geq 0\right)$. However, the general 5D geodesic contains extra terms related to the fifth dimension, and these can show up as modifications of the standard 4D equations of motion. Notably, there is an extra force in both current versions of 5D relativity [10, 11]. This acts parallel to the 4-velocity, so from the standpoint of the usual law of conservation of momentum is most naturally expressed as a variation of the rest mass of a test particle along its 4D path. In addition to these changes to the equations of motion, there are changes to the field equations of the theory. These alter the limit where there is no ordinary matter present, and are most naturally expressed as a modification of the energy density of the vacuum as measured by the cosmological 'constant', which in general is now variable.

The behaviours of *m* and $\Lambda$ in 5D relativity have recently been studied by a couple of workers [12, 13]. These analyses appear superficially different, since they concentrate on classical and quantum aspects of the problem. But both are motivated by the past discovery that the forms of *m* and $\Lambda$ depend on the choice of the extra coordinate $x^4 = l$ [14]. This is best demonstrated by applying *l*-dependent conformal factors to the metric, and means that the 4D values of *m* and $\Lambda$ appear to be gauge-dependent. The objective of the present work is to clarify this, and to bring out the connection between *m* and $\Lambda$. In order to do this, it will be necessary to recall certain results on 5D canonical metrics in the first part of Section 2 (some readers may be familiar with this subject). Then the status of $m$ is investigated, followed by $\Lambda$ in Section 3. The last section is largely discussion. The notation is standard, with $\alpha, \beta = 0,123$ and $A, B = 0,123,4$. The constants $(c, G, h)$ are made explicit at places to aid physical understanding.

2. Conservation of Momentum and $m$

In this section and the next, it is assumed that the motion in 5D is given by the geodesic equation; and that it can be solved by obtaining the metric coefficients $g_{AB} = g_{AB}(x^\gamma, l)$ from the standard 5D field equations $R_{AB} = 0$ ($A, B = 0,123,4$). The latter can always be reduced to simpler relations once a metric is specified, and they include Einstein's 4D field equations, usually with a cosmological constant $\Lambda$. This reduction from 5D to 4D is guaranteed by an old embedding theorem due to Campbell, and the consequences of the procedure are by now well known (see ref. 3 for a review). The 5D geodesic equation can be split into a 4D part and an extra part. The former can be expressed as 4D geodesic motion plus correction terms, while the latter can be expressed as an equation for the evolution of the extra coordinate.

For present purposes, it is desirable to compare the 4D motion with that expected from the law of conservation of (linear) momentum, defined as usual by the product of the mass and the 4-velocity $u^\mu \equiv dx^\mu / ds$. Globally, the momentum is expected to be affected by the curvature. Locally, however, to good accuracy it obeys $d(mu^\alpha)/ds = 0$. That is,

$$\frac{du^\mu}{ds} + \frac{u^\mu}{m}\frac{dm}{ds} = 0 \quad . \tag{1}$$

The first term here is the conventional acceleration, but the second term could harbour unconventional effects associated with the fifth dimension.

To investigate this, it is necessary to specify a metric. The full 5D metric tensor, defined via $dS^2 = g_{AB}dx^A dx^B$, has 15 independent components. Of these, 5 can be removed without compromising the generality of the metric by using the 5 available degrees of coordinate freedom to set $g_{4\alpha} \to 0$, $|g_{44}| \to 1$. Physically, this means that the electromagnetic and scalar fields are either absent or can be treated as parts of the gravitational sector of the theory, $g_{\alpha\beta} = g_{\alpha\beta}(x^\gamma, l)$ in general. With this prescription, it is possible to work out the 4D and extra parts of the 5D geodesic equation. To compare with (1), the 4D motion is relevant. And because the motion may be 5D null $(dS^2 = 0)$, and because we wish to make contact with known dynamics, it is best to use the 4D proper time as parameter, defined by $ds^2 = g_{\alpha\beta}(x^\gamma, l) dx^\alpha dx^\beta$. Then the metric and the motion are given by

$$dS^2 = g_{\alpha\beta}(x^\gamma, l) dx^\alpha dx^\beta - dl^2 \tag{2.1}$$

$$\frac{du^\mu}{ds} + \Gamma^\mu_{\beta\gamma} u^\beta u^\gamma = F^\mu \quad . \tag{2.2}$$

Here the 5D metric (2.1) is taken to have a spacelike extra dimension, but in principle it could be timelike, since the extra coordinate does not have the physical nature of a time (see below). The Christoffel symbols $\Gamma^{\mu}_{\beta\gamma}$ embody the force usually called gravitational, while $F^{\mu}$ is a force (per unit mass) which includes other terms not present in general relativity. This force can actually contain parts normal and parallel to the 4-velocity: $F^{\mu} = N^{\mu} + P^{\mu}$. However, it is the latter which is responsible for the unconventional second term in the conservation-based relation (1). It is given by

$$P^{\mu} = \left(-\frac{1}{2}\frac{\partial g_{\alpha\beta}}{\partial l}u^{\alpha}u^{\beta}\right)\frac{dl}{ds}u^{\mu} \quad . \tag{3}$$

The term in parentheses here represents the coupling between the $l$-dependent 4D metric and the 4-velocities. The force (per unit mass) is proportional to the 'velocity' in the extra dimension, measured with respect to 4D proper time. It should be noted that the metric (2.1) is invariant under $l \to -l$, so $dl/ds$ is reversible. The same applies to the dependency of $P^{\mu}$ on $u^{\mu}$. The force (3) is inertial in the Einstein / Mach sense [1], since it exists because of the relative motion of two frames of reference.

Combining (1) and (3) gives

$$\frac{1}{m}\frac{dm}{ds} = \left(\frac{1}{2}\frac{\partial g_{\alpha\beta}}{\partial l}u^{\alpha}u^{\beta}\right)\frac{dl}{ds} \quad . \tag{4}$$

The mass of a test particle is seen to be variable along its path, given that the metric is dependent on the extra coordinate and that the velocity is finite in the extra dimension. These things are given by solutions of the field equations and the equations of motion. But (4) shows also that the mass can in principle be stabilized by the background, which will usually contain vacuum energy and ordinary matter. [The energy-momentum tensor $T_{\alpha\beta}$ derived from $g_{\alpha\beta}(x^\gamma, l)$ has a standard form, given for example in ref. 3 p. 16.] However, without going into these details it is possible to obtain an expression for $m = m(l)$ from (4). In that relation, the velocity $dl/ds$ is inherently reversible, as noted above. Taking account of this by a sign choice, and introducing an appropriate fiducial value of the mass, the integration of (4) yields

$$m = m_* \exp\left[\pm \int \left(\frac{1}{2} \frac{\partial g_{\alpha\beta}}{\partial l} u^\alpha u^\beta\right) dl\right] \quad . \tag{5}$$

This is an explicit relation for the mass of a local test particle embedded in a background universe with the general metric (2.1), and as such realizes Mach's Principle.

A specific application of (5) is when the metric tensor in (2.1) is restricted to be $g_{\alpha\beta}(x^\gamma, l) = (l/L)^2 \bar{g}_{\alpha\beta}(x^\gamma \text{ only})$. This simplifies the coupling factor in (5), and also streamlines the field equation and the equations of motion (see below). It should be noted that (2)-(5) assume that the 4-velocities are normalized via $g_{\alpha\beta}(x^\gamma, l) u^\alpha u^\beta = 1$, *not* $\bar{g}_{\alpha\beta}(x^\gamma) u^\alpha u^\beta = 1$. The pertinent value of the coupling factor in (5) is thus $(1/2)(\partial g_{\alpha\beta}/\partial l) u^\alpha u^\beta = 1/l$. This does not depend on the constant length $L$ introduced just above for the consistency of physical dimensions or units. The physical meaning of this parameter will be clarified below. Here I note that the factorization of the 4D part of the 5D metric by a quadratic term in $x^4 = l$ has been widely used in STM-theory, while an alternative factorization using an exponential term has been used in M-theory [3]. The quadratic form has certain advantages, one of which is that (5) for the mass of a test particle gives a very simple result:

$$\frac{m}{m_*} = \left(\frac{l}{l_*}\right)^{\pm 1} . \qquad (6)$$

The integration constants have no special significance, while the sign choice merely reflects the reversibility of the motion in the extra dimension. The true significance of (6) is that *particle mass is related to the extra coordinate of 5D relativity.*

This striking result is for a metric with the form

$$dS^2 = (l/L)^2 \bar{g}_{\alpha\beta}(x^\gamma) dx^\alpha dx^\beta - dl^2 . \qquad (7)$$

This form is frequently called canonical, but it will be seen below that there is an important distinction to be made for metrics of this form about the nature of the spacetime coordinates. The simple result (6) depends on using coordinates $x^\gamma$ which normalize the 4-velocities in a way that includes $x^4 = l$, via the previously-noted condition $g_{\alpha\beta}(x^\gamma, l) u^\alpha u^\beta = 1$ where $g_{\alpha\beta} = (l/L)^2 \bar{g}_{\alpha\beta}(x^\gamma)$. This implies that an observer doing dynamical experiments is effectively using the spacetime specified by $g_{\alpha\beta}$, not $\bar{g}_{\alpha\beta}$. An observer unaware of the fifth dimension is really using this kind of hybrid spacetime, and can expect to find masses given by (6), which via $l = l(s)$ will in general be variable.

The algebraic result (6) has a logical physical interpretation. The kind of mass which figures in non-gravitational problems, and which is ubiquitous in quantum me-chanics, can be geometrized via the Compton wavelength $h/mc$. Alternatively, the kind of mass which figures in gravitational problems can be geometrized via the Schwarzschild radius $Gm/c^2$. Einstein's Weak Equivalence Principle (WEP) says that these kinds of mass are proportional to each other. Even so, it is convenient to label the two as $l_i$ and $l_g$ respectively, for the inertial and gravitational measures of particle mass via lengths. Then the sign choice in (6) is obviously an indicator of the appropriate choice of mass coordinate. The implication is that the WEP is a reflection of the reversibility of the motion in the fifth dimension. In a previous interpretation of 5D relativity, the interchangeability of $l_i$ and $l_g$ was ascribed to the coordinate transformation $l_i \to L^2/l_g$ [3]. But the reversibility argument is simpler. Both, it should be noted, imply

$$l_i \, l_g = \left(\frac{h}{mc}\right)\left(\frac{Gm}{c^2}\right) = \frac{Gh}{c^3} = l_{\text{Plank}}^2 \quad . \tag{8}$$

The Planck length plays a central role in the problem of how to construct a quantum theory of gravity and the cosmological-'constant' problem. But caution is needed here, because from the viewpoint of 5D relativity the lengths $l_i$ and $l_g$ are merely different ways of measuring the same thing, namely mass.

Returning to the question of metrics and mass variability, there is an alternative form of the canonical metric (7) in which the 4-velocities are normalized in a different way, namely by using the 4D metric *without* the $(l/L)^2$ prefactor taken into account. That is, the first part of the metric is written as $(l/L)^2 \tilde{g}_{\alpha\beta}(x^\gamma, l) dx^\alpha dx^\beta$, and the veloci-ties are normalized via $\tilde{g}_{\alpha\beta}(x^\gamma, l) u^\alpha u^\beta = 1$. The observer now experiences a spacetime with $ds^2 = \tilde{g}_{\alpha\beta}(x^\gamma, l) dx^\alpha dx^\beta$ which may be *l*-dependent but is decoupled from the $(l/L)^2$ prefactor. This prescription intrinsically defines a different set of coordinates from what was considered before. Much has been written about this form of the metric, so it is useful to give the full set of relations for the metric plus the components of the geodesic equation for spacetime and the extra dimension:

$$dS^2 = (l/L)^2 \tilde{g}_{\alpha\beta}(x^\gamma, l) dx^\alpha dx^\beta - dl^2 \tag{9.1}$$

$$\frac{du^\mu}{ds} + \Gamma^\mu_{\beta\gamma} u^\beta u^\gamma = F^\mu$$

$$F^\mu \equiv \left( -\tilde{g}^{\mu\alpha} + \frac{1}{2} \frac{dx^\mu}{ds} \frac{dx^\alpha}{ds} \right) \frac{dl}{ds} \frac{dx^\beta}{ds} \frac{\partial \tilde{g}_{\alpha\beta}}{\partial l} \tag{9.2}$$

$$\frac{d^2 l}{ds^2} - \frac{2}{l}\left(\frac{dl}{ds}\right)^2 + \frac{l}{L^2} = -\frac{1}{2}\left[\frac{l^2}{L^2} - \left(\frac{dl}{ds}\right)^2\right] \frac{dx^\alpha}{ds} \frac{dx^\beta}{ds} \frac{\partial \tilde{g}_{\alpha\beta}}{\partial l} \quad . \tag{9.3}$$

This prescription is general, like (2). As there, the motion in spacetime is geodesic with an extra force, or acceleration per unit mass, which is dependent on the behaviour of the extra coordinate. The second part of this extra force has the same form as (3). The evolution of the extra dimension is given by (9.3). The latter equation has a generic solution, independent of $\partial \tilde{g}_{\alpha\beta} / \partial l$, given by $l = l_* \exp(\pm s/L)$. This is also the path given directly from (9.1) for *null*-paths with $dS^2 = 0$. Since $\partial \tilde{g}_{\alpha\beta} / \partial l \neq 0$ in (9.1) is known to imply the presence of ordinary matter [3], the physical picture given by (9) is one of a universe filled with particles which are all in 5D causal contact.

The metric (9.1) is algebraically general, as noted above. But the special case where $\tilde{g}_{\alpha\beta} = \tilde{g}_{\alpha\beta}(x^\gamma \text{ only})$ is of special interest. Then there is no extra force (9.2), so the 4D motion is geodesic in the conventional sense. Accordingly, by the conservation of momentum, the mass of a particle is constant along its path. Also, by reduction of the 5D field equations to their 4D counterparts [3], there is no ordinary matter present, but there is vacuum energy with a finite cosmological constant $\Lambda = +3/L^2$. If it is desired to reverse the sign of this parameter, it can be done for canonical-type metrics by changing the extra dimension from spacelike to timelike. This changes the nature of $l$, which is now not monotonic but oscillating with $l = l_* \exp(\pm is/L)$. This is the same as the wave function of (old) quantum mechanics if $L$ measures the mass and is identified with the Compton wavelength $h/mc$ of the particle associated with the wave. In fact, the extra component of the geodesic equation now becomes the Klein-Gordon equation [13; this subject will be revisited in Section 3]. Reverting to the more common choice of a spacelike extra dimension, for the case $\partial \tilde{g}_{\alpha\beta}/\partial l = 0$ under consideration, the situation may be summed up as follows:

$$dS^2 = (l/L)^2 ds^2 - dl^2 \qquad (10.1)$$

$$ds^2 = \tilde{g}_{\alpha\beta}(x^\gamma \text{ only}) dx^\alpha dx^\beta \qquad (10.2)$$

$$\Lambda = 3/L^2 \quad , \quad m = m_* = \text{constant} \quad . \tag{10.3}$$

For obvious reasons, this is frequently called the pure-canonical case. It has a large literature, but can actually be generalized with some unexpected consequences, as will be seen in the next section. The importance of the pure-canonical case (10) is that it represents an embedding in 5D which leaves the 4D physics of general relativity intact.

3. <u>Conformal Factors and $\Lambda$</u>

5D metrics of canonical type possess by construction a flat extra dimension, with a 4D part which consists of an $x^4$-dependent prefactor multiplied onto a spacetime metric which in general depends on $x^\gamma$ and $x^4 = l$. This kind of metric can in theory describe *any* 5D situation; though if an exact solution of the field equations is found in non-canonical coordinates, the metrics may have to be brought into correspondence by applying a coordinate transformation. The latter, of course, is not in general equivalent to a conformal transformation, wherein the prefactor on the 4D part of the 5D metric is changed. In the present section, some technical results will be recalled concerning conformal transformations, in order to see what new physics emerges, particularly about the cosmological 'constant'.

The pure canonical metric (10) is algebraically restricted insofar as $\tilde{g}_{\alpha\beta} = \tilde{g}_{\alpha\beta}(x^\gamma \text{ only})$. However, its prefactor can be generalized by a shift $l \to (l - l_0)$ along the extra axis, since this leaves the last term in (10.1) unchanged. This coordinate transformation, while seemingly trivial, has a surprisingly drastic effect on $\Lambda$ [12-14]. The shifted pure-canonical metric has:

$$dS^2 = \left(\frac{l-l_0}{L}\right)^2 ds^2 - dl^2 \tag{11.1}$$

$$ds^2 = \tilde{g}_{\alpha\beta}(x^\gamma \text{ only}) dx^\alpha dx^\beta \tag{11.2}$$

$$\Lambda = \frac{3}{L^2}\left(\frac{l}{l-l_0}\right)^2 . \tag{11.3}$$

The divergence in the effective value of the 4D cosmological 'constant' for $l \to l_0$ has profound physical implications. These can be worked out given a relation $l = l(s)$, such as the one which follows from the null-path condition $(dS^2 = 0)$ mentioned previously. For a spacelike extra dimension as in (11.1), there results a cosmological model where $\Lambda > 0$ drives early inflation and later decays to a value consistent with supernova data [4-6]. For a timelike extra dimension, there results a model for a particle where $\Lambda < 0$ drives a quantum wave with the usual rule of quantization [13]. It is natural to inquire more deeply about the nature and effects of conformal transformations.

    The result (11.3) was derived from two long and slightly different algebraic manipulations of the field equations by Mashhoon and Wesson [14]. But they noted that a more direct result to $\Lambda$ lay in using properties of the 4D Ricci tensor $R_{\alpha\beta}$ and the curvature scalar $R$. This method was in effect used to verify the result (11.3) by Ponce de Leon [12]. Even so, as Ponce de Leon noted, there are certain "subtleties" in the calculations (a detailed discussion for the dedicated worker is given in ref. 15). In metrics of pure-canonical type, where the prefactor depends only on $x^4 = l$ and the spacetime tensor de-

pends only on the 4D coordinates $x^\gamma$, the conformal prefactor acts algebraically like a constant. It is a theorem, that under a constant conformal transformation, the components of $R_{\alpha\beta}$ are unchanged [16]. However, because the contraction of $R_{\alpha\beta}$ involves the conformal factor, the latter affects the value of $R$. And because in a vacuum spacetime $R = 4\Lambda$, the cosmological constant is also affected. Thus for metrics of pure-canonical type, including (10) and (11), the application of a conformal factor $\Omega^2(l)$ involves:

$$g'_{\alpha\beta} = \Omega^2 g_{\alpha\beta}, \quad R' = \Omega^{-2} R, \quad \Lambda' = \Omega^{-2} \Lambda \quad . \tag{12}$$

To go from (10.1) to (11.1), the appropriate factor is given by $\Omega^2 = (l - l_0)^2 / l^2$. This in effect 'removes' the original $l$-factor of the simple canonical metric and 'replaces' it by the shifted factor $(l - l_0)$. More complicated changes of the conformal factor can be handled by applying (12) to (11).

In retrospect, it is not too surprising that $\Lambda$ transforms via a conformal factor applied to the 4D part of a 5D metric. It is introduced in Einstein's equations as $\Lambda g_{\alpha\beta}$, so must respond as in (12) to conformal changes of the metric tensor. What is more difficult to appreciate is that a 4D quantity (such as $\Lambda$) can depend on $x^4 = l$ and change under a 5D coordinate transformation which includes $l$. But this is also clear in retrospect. The 5D theory is covariant in 5D, not 4D. The field equations $R_{AB} = 0$ are manifestly covariant; and the equations of motion are given by the components of the 5D geodesic equation, which are an embodiment of the invariant variational statement $\delta\left[\int dS\right] = 0$ (most conveniently taken around the null state). It has been pointed out by several workers that the 5D group of coordinate transformations is broader than the 4D group. That is,

$$x^A \to \bar{x}^A(x^B) \quad , \quad x^\alpha \to \bar{x}^\alpha(x^\beta) \tag{13}$$

are not equivalent, so a change in the 5D gauge can now lead to a change in the form of a 4D quantity. In other words, the cosmological 'constant', while introduced as a true constant in 4D, is gauge-dependent in 5D.

Similar comments apply to the rest mass $m$ of a test particle. That is why the 4D equations of motion, such as (2.2) and (9.2), have an extra force (per unit mass) $F^\mu$. Those relations, like in Einstein's theory, actually involve accelerations. These are connected to forces, which figure in Newton's mechanics, by the mass $m$. In order to preserve the conservation of momentum, $m$ in general varies in 5D, as stated for example in (4) which shows that $m(l)$ is gauge-dependent. Another way of seeing this is to look at the functional form of the extra 'force' which appears in (3) and (9.2). The essential factor is $(\partial g_{\alpha\beta}/\partial l)u^\alpha u^\beta (dl/ds)$, which can be rewritten with $(\partial g_{\alpha\beta}/\partial l)dl = d(g_{\alpha\beta})_l$. This is the change in the metric tensor of spacetime caused by moving away from the original hypersurface to a new one, along the direction of $x^4 = l$, a change which to preserve momentum has to involve a change in $m(l)$. Further, the change in $m$ is expected to be of the same kind as that for $\Lambda$, namely by a conformal factor. This is obvious when it is recalled that the action for a massive particle is commonly taken to be $I = \int mcds$, where the mass ought to be kept inside the integral in case it is variable. This is the quantity which leads to quantization in terms of Plank's constant $h$ (the two things have the same physical dimensions or units, so $I$ can be written in dimensionless form using $l_i = h/mc$ of Section 2). Let us assume that we wish to have $I$ invariant under conformal transformations, in the same way as for Einstein's equations, where $\Lambda g_{\alpha\beta} \sim \Omega^{-2} \cdot \Omega^2 =$ constant by (12). The quantity $(mds)^2$ is proportional to $g_{\alpha\beta}$, so if the latter scales by $\Omega^2$ then we require that $m^2$ scales as $\Omega^{-2}$. That is, $m^2$ is changed under a conformal transformation in the same manner as $\Lambda$. Therefore:

$$g'_{\alpha\beta} = \Omega^2 g_{\alpha\beta}, \quad m' = \Omega^{-1} m \quad . \tag{14}$$

This complements (12). The two relations together complete the transformations which preserve the quantities of 4D general relativity and quantum mechanics when the metrics are extended to 5D and have canonical form.

It should be noted that the preceding paragraph effectively chooses to geometrize mass by the length $l_i = h/mc$, rather than the alternative $l_g = Gm/c^2$. These parameters were introduced in Section 2, and their product related to the Planck length in equation (8). An equivalent nomenclature is in terms of atomic and gravitational units of measure, as employed in older work on 4D scale-invariant field theory. While each choice of parameters is valid in its own domain, the joining of quantum and gravitational physics in 5D, via the use of conformal factors, appears to prefer the former one. It may be confirmed that with the noted choice, (14) in combination with the three metrics (2.1) / (7), (9.1) and (10.1) give the same expressions for $m = m(l)$ as found by Ponce de Leon [12]. The simplest of these cases is the pure-canonical metric (10.1), which was shown above to give constant particle masses. This is, of course, in agreement with laboratory physics, and indicates that for short distances and times, the pure-canonical coordinates are the ones employed by local physics. By (10.1), the appropriate identification is $L = h/mc$, which gives back the standard action noted above, and agrees with the application to wave mechanics [13]. However, it is quite feasible that because of the large-scale evolution of the universe, cosmology may be best described by another form of the canonical metric, such as (9.1). That metric implies variable particle masses. To see this, we add a shift for generality to the $l$-measure of (9.1), so the appropriate conformal factor is the same one considered above in connection with $\Lambda$ of (11.3), namely $\Omega = (l - l_0)/l$. Then compared to the constant $m_*$ of the pure-canonical case (10), metric (9.1) with (14) gives

$$m = m_* \left| \frac{l}{l - l_0} \right| \quad . \tag{15}$$

Here $l = l(s)$ evolves as the universe expands, and is given by solving the extra component of the geodesic equation (9.3), or directly for 5D null-paths by the metric (9.1). If $l$ evolves from relatively small values to large ones, then the mass of a particle is negligible just after the big bang, and tends to a constant at late epochs. This kind of model is astrophysically appealing, because it accounts for the thermalization of background radiation via divergent scattering cross-sections at early times, but agrees with the constancy of masses at late times. However, it needs detailed investigation.

4. Conclusion and Discussion

Five-dimensional relativity occupies a place between Einstein's theory and higher-dimensional theories which might explain all of the interactions of particles.

However, 4D physics may appear to be gauge-dependent, because 5D admits a broader group of coordinate transformations, as well as conformal transformations. The latter are relevant, because the action $mds$ and the cosmological term $\Lambda g_{\alpha\beta}$ are both proportional to the 4D metric tensor; and when this changes, the preservation of the noted quantities requires in general that $m$ and $\Lambda$ change: $m = m(l)$, $\Lambda = \Lambda(l)$. There are subtle problems of interpretation connected with these coordinate and conformal transformations, which the present account aims to clarify. To bring 5D relativity into closer correspondence with conventional mechanics, the electromagnetic and scalar potentials have been suppressed while leaving the 5D metric otherwise general, as in (2) and (9). By contrast, (7) and (10) are special, but differ in how the 4-velocities are normalized. In all four cases, the standard 4D proper time $s$ is used as the parameter for the dynamics. This choice is consistent with the postulate that while massive particles move on timelike paths in 4D $(ds^2 > 0)$, they can move on null paths in 5D $(dS^2 = 0)$. These photon-like paths agree with supersymmetry in 5D, though this is broken by the gauge-dependent masses and the cosmological 'constant' in 4D. Expressions for these quantities in certain important cases have been given. Generally, they scale under a conformal transformation $\Omega$ of the 4D metric tensor via $\Lambda' = \Omega^{-2}\Lambda$ and $m' = \Omega^{-1}m$, as in (12) and (14). There is therefore a relationship between the mass of a particle and the local value of the cosmological 'constant'. In its simplest form, this is $\Lambda \sim m^2$, a relation which can in principle be tested.

It has become common practice, especially in 4D Friedmann-Robertson-Walker cosmology, to express $\Lambda$ as the energy density of the vacuum via the equation of state $\rho_v = -p_v/c^2 = \Lambda c^2/8\pi G$. Likewise, in the Schwarzschild-deSitter solution the potential can be written in terms of $(2G/c^2 r)(M_* + M_v)$ where $M_*$ is the mass of the central star and $M_v = (4\pi r^3/3)\rho_v$ is the 'mass' of the surrounding vacuum out to radius $r$. This may be convenient, but can be misleading: the coupling constant between geometry and matter in general relativity is $8\pi G/c^2$, which exactly cancels the physical constants in the energy-momentum tensor for the density of the vacuum $\Lambda c^2/8\pi G$, leaving the term $\Lambda g_{\alpha\beta}$ as it was in the original form of Einstein's equations. Instead of hiding the $\Lambda$-term, it is more instructive to regard it as a kind of gauge term for Einstein's equations, analogous to the one in Maxwell's equations. This is fully consistent with the 5D view, where $\Lambda = \Lambda(l)$ depends generally on the choice of coordinates and especially $x^4 = l$. It is also consistent with the view, expressed previously, that Mach's Principle is realized in 5D.

For illustration, consider a simple Machian thought experiment. Take a single particle and place it in a previously empty universe which is defined to have zero global energy. Then what is the mass of the particle? From the 5D Machian viewpoint, the particle is now surrounded by fields which permeate space. In general, these would be the gravitational field (whose energy cannot be localized), a scalar field (if it is not suppressed by a choice of coordinates), and a field corresponding to the cosmological 'constant' (which in 5D is dependent on the coordinate $x^4 = l$). Assuming that the total energy of the universe is conserved and zero, the conclusion must be that the positive rest mass of the particle is balanced by the negative energy of the fields. This implies that there has to be a relationship between $m$ and $\Lambda$. Such a relationship may be complicated when the universe has many particles; but the connection between *m* and Λ derived in the present account for 5D is consistent with what Mach (and Einstein) thought about mechanics in 4D [1]. Present results are also consistent with technical proofs by various workers that 4D FRW models have global measures of energy which are actually zero [17]. Present results agree as well with the discovery, made repeatedly by different workers [18], that 4D FRW models can be embedded in *flat* 5D manifolds.

Ways in which the preceding account may be extended are numerous and mainly technical. However, the overall view which emerges from recent work on 5D relativity is that the universe may in some sense be empty, at least when considered from a higher-dimensional perspective.

Acknowledgements

Thanks for past comments go to B. Mashhoon, the late H. Liu and J. Ponce de Leon. This work was partially supported by NSERC.

studies, the 5D and 4D intervals are assumed to be related by $dS = fds$, where $f$ is a function of coordinates. Even though many useful results are obtained in these two studies, the use of the $f$-factor is in some ways unfortunate. For example, it does not allow of the smooth recovery of results for the null case $(dS^2 = 0)$, which has to be treated separately; and it depends on how the 4-velocities are normalized. For $dS^2 = (l/L)^2 \left[ g_{\alpha\beta}(x^\gamma,l) dx^\alpha dx^\beta \pm dl^2 \right]$ it is given by $f^2 = \left[ (l/L)^2 \pm (dl/ds)^2 \right]$, assuming the 4-velocities $u^\alpha \equiv dx^\alpha / ds$ are normalized via $g_{\alpha\beta}(x^\gamma,l) u^\alpha u^\beta = 1$, in which case the extra component of the 5D geodesic equation agrees with that in refs. 10 and 13. It should also be noted that some accounts presume that the scalar-field function $\Phi$ is real, and that the 5D signature is $(+----)$. However, if contact is to be made with the Higgs field of ref. 7, $\Phi$ should be a complex function, leading to the possibility that in certain regions the signature is $(+---+)$. This is why ref. 12 lacks the quantum phenomena discussed in ref. 13. Of course, observations limited to 4D spacetime cannot directly reveal $g_{44} = \pm \Phi^2$. And the structure of the 5D field equations suggests compelling coordinate transformations which render $g_{44} = \pm 1$. (These are implied in ref. 10 and spelled out in: P.S. Wesson, ArXiv: gr-qc/1011.0214, 2010.) In classical terms, a change in the sign of $g_{44}$ can be accounted for by an horizon in the fifth dimension, of the sort which may occur in the class of 5D exact solutions known as solitons (for a review of these objects, see: P.S. Wesson, ArXiv: gr-qc/1104.3244, 2011). It should further be noted that some accounts ignore the reversibility of the motion in the extra dimension. This is not trivial, because a change in the sign of $dl/ds$ reverses the sign of the so-called fifth force of 5D relativity. For the canonical metric, the application of the conservation of momentum means that the sign choice for $dl/ds$ affects the physical identification of the particle rest mass. Hence the appearance of the mass measures $h/mc$ and $Gm/c^2$ in Section 2 of the main text.

[16] J.L. Synge, Relativity: The General Theory, North Holland, Amsterdam (1960), 318. R. Wald, General Relativity, Un. Chicago Press, Chicago (1984) 445. P.S. Wesson, Space-Time-Matter, 2nd ed., World Scientific, Singapore (2007) 242.

[17] V. Faraoni, F.I. Cooperstock, Astrophys. J. 587 (2003) 483, and references therein.

[18] M. Lachieze-Rey, Astron, Astrophys. 364 (2000) 894. I.E. Gulamov, M.N. Smolyakov, Gen. Rel. Grav., in press (2012), and references therein.